\documentstyle[prl,twocolumn,aps,epsf]{revtex}
\begin{document}
\twocolumn[\hsize\textwidth\columnwidth\hsize\csname@twocolumnfalse%
\endcsname

\title{Triggering synchronized oscillations through
arbitrarily weak diversity in close-to-threshold excitable media}

\author{C. Degli Esposti Boschi and E. Louis}
\address{Departamento de F{\'\i}sica Aplicada and Unidad Asociada of the 
Consejo Superior de Investigaciones Cient{\'\i}ficas,   \\
Universidad de Alicante,  Apartado 99, E-03080 Alicante, Spain.}

\author{G. Ortega}
\address{Centro de Estudios e Investigaciones,
Universidad Nacional de Quilmes,\\ R. S. Pe\~na 180, 1876, Bernal, Argentine.}

\date{\today}
\maketitle

\begin{abstract}
It is shown that an arbitrarily weak (frozen) heterogeneity can induce global 
synchronized oscillations in excitable media close to threshold.
The work is carried out on networks of coupled van 
der Pol-FitzHugh-Nagumo oscillators.  The result is shown to be robust 
against the presence of internal dynamical noise.
\end{abstract}

]

\narrowtext
Emergent synchronized oscillations is a key subject in a variety of fields
ranging from physics to biology or medical sciences. In the last
few years several papers have been published concerned with the possibility
of triggering global oscillatory behavior through  heterogeneity
and/or internal (dynamical) noise
in the excitable medium \cite{SR93,Sh94,Ca00,HZ00,DS00,DS01,AP01}. Although
these works are a significant step towards the understanding
of emergent oscillatory behavior, many  points remain unclear. For
instance, in  \cite{Ca00} a model was proposed to investigate
emergent oscillations in pancreatic $\beta$ cells \cite{DM68}, in which half
of the elements was in the silent phase while the other half
was continuously active. Although the approach gave clean globally
synchronized oscillations, it is doubtful whether such  a {\it symmetric}
arrangement may have any biological meaning. The combined effects of 
diversity \cite{SR93,Sh94,Ca00}
and internal dynamical noise \cite{DS00,PK97,RP99} have also been 
investigated \cite{HZ00,DS01,NS99,SF01}.
While, a common conclusion seems to be the significant role
of dynamical noise in triggering global oscillations,
a key point such as the size dependence of the results was
not investigated in detail. Here we start from Cartwright approach
\cite{Ca00} and explore the possibility of synchronization as
a function of the amount of diversity (fraction of diverse elements). 
We show that, as the system approaches threshold for
oscillatory behavior, the number of 
diverse elements required to trigger global oscillations becomes 
arbitrarily small.
This is particularly appealing from a biological point of view as
the possibility of having a small number of cells different from
the rest is always there. We also show that these results are
not significantly affected by internal dynamical noise.

We base our analysis upon the van der Pol--FitzHugh--Nagumo equations 
\cite{vv28,Fi60,Fi61,NA62} that, as discussed in \cite{Ca00}, are
an adequate mathematical description of the circuit model (which
involves a capacitor, a nonlinear resistance accross the capacitor,
an inductance and a resistance) commonly
used to represent a physiological excitable medium. Including
first--nearest-neighbors coupling between the elements in the 
network \cite{Ca00,Wi91,Wi67} the equations are written as,

\begin{mathletters}
\begin{equation}
{\dot \psi_i}=\gamma(\phi_i-\psi_i^3/3+\psi_i)
\end{equation} 
\begin{equation}
{\dot \phi_i}=-\gamma^{-1}[\psi_i+(\nu_i+\nu_0\eta(t))+\beta\phi_i]+
\kappa\sum_{j=1}^{N}(\psi_j-\psi_i)\;.
\end{equation} 
\end{mathletters}

\noindent where $N$ is the number of elements in the network. 
All constants and variables are dimensionless. 
Variables $\psi$ and $\phi$ are proportional to the potential
accross the nonlinear resistance (cell membrane) and the current through
the supply, respectively. The subindex $i$ indicates an element in
the network. The constant $\nu$ is proportional to the potential
supplied, $\beta$ to the (membrane) resistance and $\gamma$ to the square root of
the quotient inductance/capacitance. 
The coupling between the elements in the
network is accounted for by $\kappa$ (see \cite{Ca00} for a thorough 
discussion). In the model we allow constant $\nu$ to be different on 
each element $i$ of the network
and to fluctuate dynamically ($\eta(t)$ is a gaussian noise and $\nu_0$
a constant).

In order to quantify the emergence of oscillatory behavior we calculate
the spatiotemporal average of variable $\psi$, namely,

\begin{equation}
\sigma_o=\sqrt{\frac{1}{N(t_f-t_i)}\sum_{j=1}^{N}\sum_{t=t_i}^{t_f}
\left [\psi^2_j(t)-<\psi_j(t)>^2\right ]}
%\label{def_sigma_o}
\end{equation} 

\noindent where $<\psi_j(t)>$ represents the temporal average of the
$j$-th potential.
The initial time $t_i$ is chosen so that the contribution
of transients to the average is minimized (note that for $\nu$
near threshold the transient can be very long, see below) 
while $t_f$ is taken to cover a sufficiently
large number of periods of the system in its oscillatory phase 
(for the values of the parameters
given below the internal period varies around 10). 

Synchronization was in its turn evaluated by calculating the following 
average,

\begin{equation}
\sigma_s=\sqrt{\frac{1}{(N-1)(t_f-t_i)}\sum_{j=2}^{N}\sum_{t=t_i}^{t_f}
\left [\psi_j(t)-\psi_1(t)\right ]^2}
%\label{def_sigma_s}
\end{equation} 

\noindent where site "1" was randomly chosen. We could have 
extended the sum to all pairs of elements $<ij>$ but this would
have prohibitively increased computation time in large networks.
Note that using $\sigma_s$ to test synchronization is 
far more demanding than most tests used in previous analyses. 
In Eqs. (2) and (3) the
discrete sum in $t$ clearly accounts for (numerical) time
averages.

In the following we take $\beta$=0.5 and $\gamma$=2 and vary the
remaining parameters.  In particular we investigate, for a given
$\nu$, how $\sigma_o$ and $\sigma_s$ vary with the fraction $x$ of elements
with $-\nu$ (hereafter referred to as impurity elements or, simply, 
impurities) distributed randomly in the network. In the absence of both 
internal noise ($\nu_0$=0) 
and coupling ($\kappa$=0), oscillatory behavior occurs for 
$|\nu|  < \nu_c=0.60412$ (for the values of $\beta$ and $\gamma$ chosen here,
see \cite{Ca00}). Calculations were carried out on $L \times L$
clusters ($L=10-40$) with periodic boundary conditions and
an integration step $\Delta t =0.002$.

In Figure 1 we plot the spatial average of $\psi$ for  $10 \times 10$ 
networks with
$\nu=0.62$ on all elements but two that have $\nu=-0.62$, with
and without coupling (in the former case $\kappa=0.5$). 
First we note that, as remarked above, the stationary state in the 
uncoupled case is only reached after rather long times ($t > 100$). 
Instead, in the coupled case the transient is
very short and the system soon shows a coherent oscillatory behavior.
Importantly, the existence of emergent oscillations do not depend
on the actual location of impurity elements. 
In Figure 2 we show the average $\psi$ 
over the network and over five
realizations of quenched disorder 
(different spatial configuration of impurities).
As the period of oscillation is weakly dependent on the
location of impurity elements,
the resulting pattern is a typical sum of oscillators 
with slightly different periods. It is interesting
to note that even in the case that the two impurities lie
at neighboring sites, global persistent oscillations emerge 
upon coupling. The results of Figure 1
are truly remarkable as global oscillations are promoted by a very
weak diversity (2\% in this case). Some characteristics of
this central result are discussed in detail hereafter. 

Figures 3 and 4 show the parameters that
characterize the emergence of oscillation and synchronization 
($\sigma_o$ and $\sigma_s$), 
versus the fraction $x$ of impurities, for $\nu$= 1.0 and 0.61, {\it i.e.}, 
far and close to threshold, respectively. 
The results correspond to networks of linear size $L$=20 and 40 with coupling 
constant $\kappa$=2 and 8 for $L$=20 and 40, respectively.  
This choice was motivated by the scaling argument of Ref. \cite{DS01}
(see also \cite{NS99} and the discussion below), according to which
one obtains solutions with similar properties in
two systems of linear size $L$ and $a L$ if the
diffusive coupling constant of the latter is increased by
a factor $a^2$ (apart from border effects).
Averages were taken over five
realizations (some checks with up to twenty realizations led to similar
results) and in the time range $t$=200--600. First we discuss
the results without dynamical noise.
The critical impurity fraction $x_c$ (value of $x$ at which $\sigma_o$ 
steeply increases)  for $\nu$=0.61 and
1 approximately lies at $x_c\approx$0.006 and 0.2, respectively,
the results being almost independent of size, particularly in the 
former case, although the sharpness of the transition to the oscillatory 
phase increases with the size of the network, as can be noted in Figure 4.
On the other hand, $x_c$ shows no dependence on the coupling
constant $\kappa$, as indicated by the results of Figure 3 and 4 and
other data not shown in the Figures (this is so once $\kappa$ is beyond
a critical value, see \cite{Ca00}). In fact, $x_c$ can be derived,
within a more than reasonable approximation, from a simple
mean field approach, according to which the onset will take place
when $<\nu>=(1-2x)\nu$ equals $\nu_c$. This leads to $x_c=0.5(1-\nu_c/\nu)$,
which gives $x_c$=0.006 and 0.198 for $\nu$=0.61 and 1.0, quite compatible
with the numerical results of Figures 3 and 4.
Note that the parameter that characterize synchronization $\sigma_s$ is
significantly smaller for $\nu=0.61$ \cite{note1}. A plausible explanation for 
this behavior is that as for $\nu=1.0$ the transition occurs at much larger
impurity concentrations, clustering is more probable, increasing the
difficulty of synchronizing the whole system. 

Dynamical noise does not qualitatively change 
the results discussed above. Figure 3 shows results
for $\nu=0.61$ and $\nu_0=1$ \cite{note2}.
The most noticeable (quantitative) changes are: 
i) At $x$=0,  $\sigma_o$ 
is  higher than in the absence of dynamical noise, although
it is still not sufficiently large so as to consider the system being in its
oscillatory phase, ii) consequently, the transition is less sharp,
and, iii) synchronization is decreased (larger $\sigma_s$).
These results are in apparent contradiction with several
analyses which indicate that  dynamical noise increases 
oscillation and synchronization \cite{HZ00,DS01,NS99}.
However, this may be well due to the non-optimal noise
level which is required for coherent resonance and to
the more severe measures of oscillation and synchronization
that we have adopted. 
Morever, we note that
in those studies nothing was said about whether the effect survives
as the size of the system increases. Preliminar results 
indicate that in fact it does not, in line with the rather small increase
of $\sigma_s$ that dynamical noise promotes near $x$=0 and the decrease
in synchronization. In
any case the main conclusion of this analysis is that dynamical noise
does not modify the previous result, that is, the dramatic
effect that a small number of impurities has in systems near threshold.

A final point concerns the effect of the coupling constant. Results
for networks of $L$=20 and 40, $\nu$=0.61 and $x$=0.02, are shown 
in Figure 5. It is noted that
$\sigma_o$ reaches its maximum (constant) value for a coupling constant
that is significantly lower for the smaller network. In fact this occurs
at $\kappa\approx$0.45 and 1.7 for $L$=20 and 40, respectively. This
is in accordance with the expected behavior (derived from
the diffusive character of the coupling term) discussed above. 
The results for the synchronization parameter are similar: to reach
the same small levels of $\sigma_s$ (less
than 0.5, say) the coupling constant in $L$=40 should
be 4 times larger. Despite
the usefulness of the scaling trick in numerical
computations, one should recall that in realistic systems
the coupling between elements
is typically intensive, that is independent of the system's size,
and it is determined by intrinsic properties.
Generally our simulations show that, when the
coupling constant is kept fixed, the emergent oscillations
and the degree of synchronization are less and less pronounced
as the number of consituents is increased. This worsening occurs
either when these effects are induced solely by noise and
when they are triggered by diversity, as discussed here.
As far as experimental results are concerned, the basic
point for the synchronous behavior to be observed is
the strength of the effective coupling constant with
respect to the number of elements. Another important
feature is the type of interaction. Indeed, it is possible
that non-diffusive couplings may lead more efficient mechanisms
of synchronization.

Summarizing, here we have discussed the possibility of triggering
global oscillations in close-to-threshold excitable media through
an arbitrarily weak heterogeneity. The work was carried out by
assuming the existence of two possible types of elements in the network, 
one silent and another continuously active. The results clearly indicate
that when the system is near threshold, global (synchronized) oscillations
emerge for a small number of diverse elements. Dynamical noise does
not alter this conclusion, although it may introduce some significant
changes such as a decrease in synchronization.

\acknowledgements
We are grateful to E. Andreu and J.V. S\'anchez-Andr\'es for
many useful comments and suggestions.
This work was supported by the Spanish "Comisi\'on Interministerial
de Ciencia y Tecnolog{\'\i}a" through grants PB96--0085 and 1FD97--1358 
and by the European Commission through the project TMR Network--Fractals 
c.n. FMRXCT980183.  GO is thankful to  the Universidad de Alicante for 
partial financial support.

\begin{figure}
\begin{picture}(236,200) (5,-15)
\epsfbox{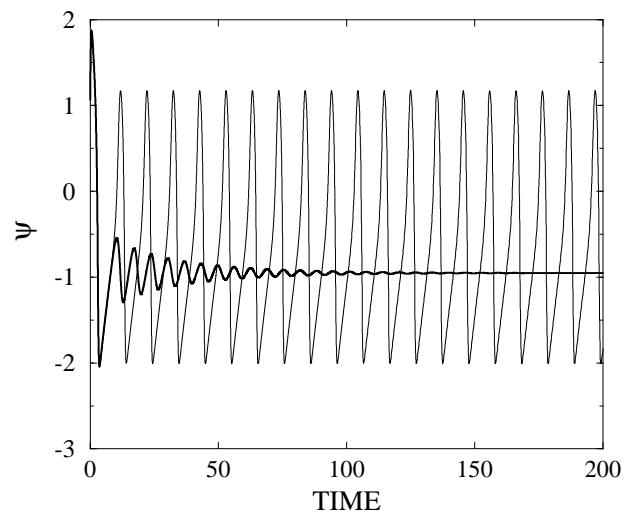}
\end{picture}
\caption{
Variable $\psi$ as a function of time in a $10 \times 10$ network
of van der Pol-FitzHugh-Nagumo elements with $\beta =0.5$, $\gamma=2$
and $\nu=0.62$ for all elements but two with $\nu=-0.62$ located
at random in the network. For those values of $\beta$ and $\gamma$ the 
threshold for 
oscillatory behavior in an isolated element occurs at $\nu{_c}=0.60412$. The
results correspond to no coupling between the elements $\kappa=0.0$ (thick 
continuous line) and for $\kappa=0.5$ (thin line). In the latter
case the average of $\psi$ over the whole network is shown.}
\label{fig.c1}
\end{figure}

%\newpage
\begin{figure}
\begin{picture}(236,220) (5,-15)
\epsfbox{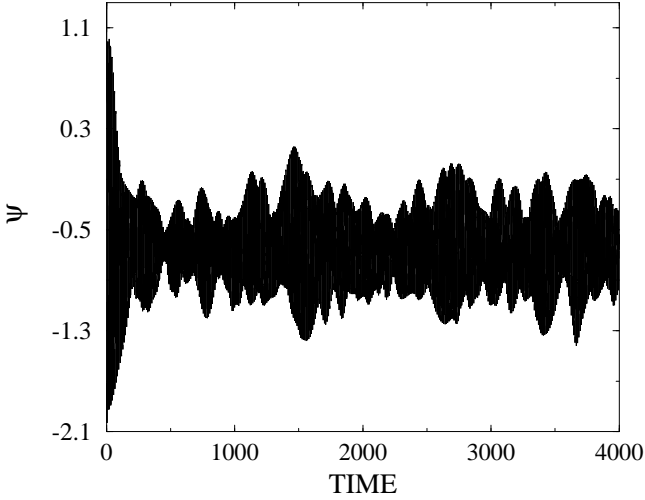}
\end{picture}
\caption{
Variable $\psi$ as a function of time in networks with the same
parameters of Fig. 1 and $\kappa=0.5$. The
result represents an average over the whole network and over five realizations 
(each corresponding to a random spatial distribution of the two elements with
$\nu=-0.62$).}
\label{fig.c2}
\end{figure}

\begin{figure}
\begin{picture}(236,220) (5,-15)
\epsfbox{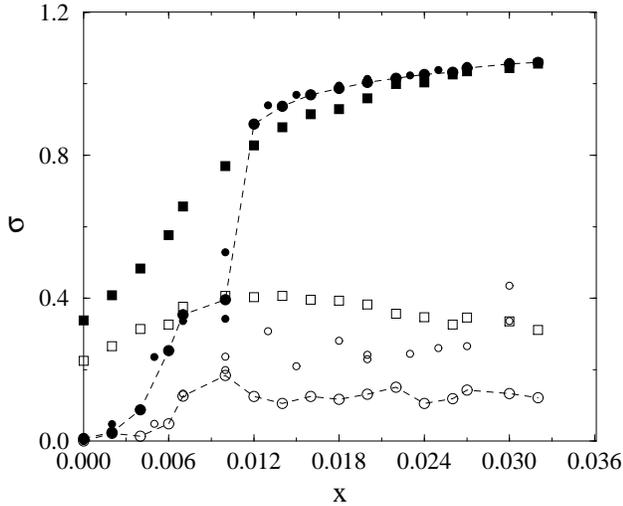}
\end{picture}
\caption{
Filled symbols: Parameter used to quantify the emergence of oscillatory 
behavior, as defined in Eq. (2), in a heterogeneous excitable
media described by Eq. (1) with $\nu=0.61$ versus the fraction of 
elements $x$ in the network with $\nu=-0.61$ (averages were done 
over 5 realizations of the disordered network). 
The numerical results correspond to networks of size $20 \times 20$ 
and $40 \times 40$ (symbol size proportional
to the linear size of the network). The rest of the parameters in the van 
der Pol-FitzHugh-Nagumo medium are: $\beta=0.5$, $\gamma=2$,
and values of $\kappa$ discussed in the text. 
Empty symbols: Same for the parameter used to quantify synchronization,
as defined in Eq. (3). Circles: without dynamical noise.
Squares: with dynamical noise ($\nu_0=1)$. The lines are guides to the eye.}
\label{fig.c3}
\end{figure}

\begin{figure}
\begin{picture}(236,220) (-5,-15)
\epsfbox{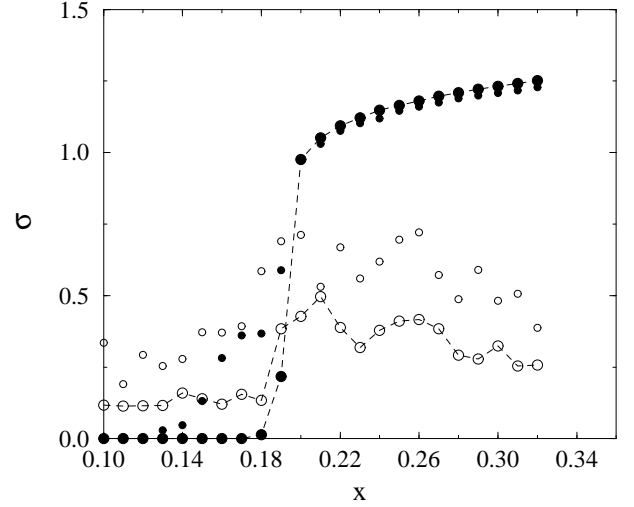}
\end{picture}
\caption{
Same as Figure 3 for $\nu=\pm 1.0$. Only results without dynamical noise 
are shown.}
\label{fig.c4}
\end{figure}

\begin{figure}
\begin{picture}(236,220) (-5,-15)
\epsfbox{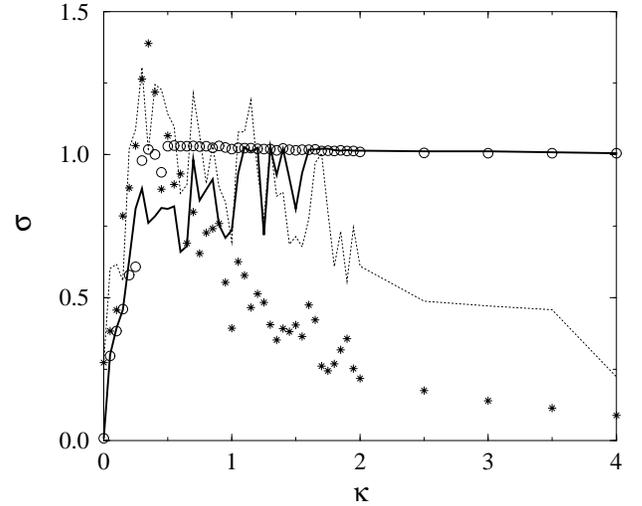}
\end{picture}
\caption{
Parameters used to quantify the emergence of oscillatory behavior 
($\sigma_o$) and synchronization ($\sigma_s$) versus coupling $\kappa$. The
results correspond to a fraction of impurities of 0.02, $\nu=0.61$,
$\beta$=0.5, $\gamma=2$ and bidimensional networks of linear
size 20 ($\sigma_o$ empty circles and $\sigma_s$ stars) and
40 ($\sigma_o$ thick continuous line and $\sigma_s$ dotted line). 
No dynamical noise was included in the calculation. Averages were done over
5 realizations of the disordered network.}
\label{fig.c5}
\end{figure}


\begin{references}
\bibitem{note1}
In absence of synchronization the parameter $\sigma_s$ is 
always above 1 (see Fig. 5).
\end{references}
\end{document}